\documentstyle[epsfig]{mn}

\title[A Dimensional study of Disk Galaxies]{A Dimensional Study of Disk Galaxies}
\author[X. Hernandez and B. Cervantes-Sodi]
{X. Hernandez$^{1}$, B. Cervantes-Sodi${^1}$  \\
$^1$ Instituto de Astronom\'\i a, Universidad Nacional Aut\'onoma de
M\'exico, A.P. 70-264, 04510 M\'exico, D.F. \\
}

\date{\today}

\begin{document}
\maketitle

\begin{abstract}
We present a highly simplified model of the dynamical structure of a disk galaxy where only
two parameters fully determine the solution, mass and angular momentum. We show through
simple physical scalings that once the mass has been fixed, the angular momentum parameter
$\lambda$ is expected to regulate such critical galactic disk properties as
colour, thickness of the disk and disk to bulge ratio. It is hence expected to
be the determinant physical ingredient resulting in a given Hubble type. A simple
analytic estimate of $\lambda$ for an observed system is provided. An explicit comparison of
the distribution of several galactic parameters against both Hubble type and $\lambda$
is performed using observed galaxies. Both such distributions exhibit highly
similar characteristics for all galactic properties studied, suggesting $\lambda$
as a physically motivated classification parameter for disk galaxies.

\end{abstract}


\section{Introduction}

The first thing an astronomer wants to know about a galaxy is its morphological type, typically
expressed through the Hubble classification scheme, introduced about 80 years ago 
(Hubble 1926, 1936). Some modifications have been introduced to the details of plan over the years
(e.g. de Vaucouleurs 1959, Sandage 1961, Kormendy 1979), although the basics 
have remained relatively unchanged. The continued usage of 
the Hubble classification owes its success to the fact that on hearing the Hubble entry of a galaxy,
one immediately forms an image of the type of galaxy that is being talked about. A wide variety of physical
features of galaxies show good monotonic correlations with Hubble type, despite the presence of significant
overlap and dispersion. To mention only a few, total magnitudes decrease towards later types (e.g. Ellis
et al 2005) and colours become bluer while gas fractions diminish from early to late types (e.g. Roberts 
\& Haynes 1994). Bulge magnitudes and bulge to disk ratios decrease (e.g. Pahre et al. 2004) 
and disks become thinner (e.g. de Grijs 1998, Kregel et al. 2002) 
in going to later types, while the relevance and structure of spiral arms also show marked trends,
with decreasing pitch angles, prominence and coherence when one reaches the earlier Hubble types
(e.g. Ma 2002, Grosbol 2004). Even the functional shape of galactic bulges correlates with
types, e.g. Graham (2001).

However, galactic type classification schemes also suffer from several shortcomings. First, the several
diagnostics which enter into the assignment of the type, appearance of the spiral pattern, bulge to
disk ratios, colour, etc., do so in a fundamentally subjective manner e.g. Adams \& Woolley (1994). 
One has to take the images to
an expert who will somehow integrate various aspects of the galaxy in question to arrive at the type
parameter. That different authors generally agree, together with the correlations between type
and various galactic parameters is indicative of a solid physical substrate to galaxy classification
schemes, the particular nature of which however, has remained elusive. Second, type parameters
are essentially qualitative (e.g. Lotz et al. 2004 ), 
which make it difficult to relate type to definitive quantitative
aspects of a galactic system, and bring into question the validity of any statistical mathematical
analysis performed on galactic populations based on type. 

Finally, as has been recognized by many
authors, the type given to a galaxy is highly dependent on the information one has regarding the system.
This stems from the inputs that determine the galactic type, the relevance of the bulge, typically
redder than the disk, means that when observing at longer rest frame bands galaxies shift to 
earlier types, an effect compounded with the high wavelength sensitivity of the bright HII regions
which define the structure and morphology of the spiral arms (e.g. Kuchinsky et al. 2000, Grosbol et al. 2004). 
This last series
of effects is of particular concern when attempting to compare galactic morphologies and types at low
and high redshifts, for example, Labbe et al. (2003) have shown that when observing the Hubble Deep Field
 galaxies in more usual
rest frame bands, many of the reported highly disturbed merger morphologies sometimes go away, to 
reveal essentially standard disks. Similarly, projection effects become important, as the structure
of the arm pattern completely disappears in edge-on disks, eliminating one of the diagnostics relevant to the
assignment of type.

In summary, certain discomfort sometimes appears regarding the use of morphological type as the principal
tool for galaxy classification, in connection to it being somewhat subjective, intrinsically
qualitative, and relative to projection and observation wavelength effects.

Several attempts at improving on these problems have appeared over the years, for example,
extensive work has been done in constructing an objective spectral classification scheme for galaxies.
These works have concentrated sometimes on the shape of the continuum (e.g. Connolly 1995), others on
the relevance of spectral emission and absorption features, as in Zaritsky (1995). These approaches 
essentially quantify properties of the underlying stellar populations, mainly yielding an 
objective description
of the relative importance of old and young components. The good correlations observed between spectral
classifications and Hubble type, indicate that indeed the status of a galaxies stellar population
varies monotonically along the Hubble sequence, although it might be a consequence of a more fundamental
dynamical/structural physics determining type, and not the defining causal determinant of type itself.
A difficulty in applying schemes of the above type massively will of course remain, in connection to the
very telescope-time intensive nature of spectroscopy, which increases with the level of detail required
e.g. in going from average shapes of the continuum to the details of spectral lines. 

A recent related
development is the work of Park \& Choi (2005) who have showed that the colour-colour gradient plane is split
into regions corresponding to distinct galactic types. This last is particularily usefull for large samples of 
galaxies, indeed, it was developed and tested using the Sloan Digital Sky Survey. Again, the physical
origin of this separation by types appers to be a characterization of the average age and locality of
the star formation history of a galaxy, which in turn would have to be traced back to some more fundamental
dynamical/formation physics behind type.

A different approach has been the use of neural networks which are trained to reproduce the subjective
process through which experts integrate features of a galaxies image to determine its type. This has the
advantage of furnishing a software which can automatically assign galactic types, without the 
direct intervention of a person e.g. Adams \& Wooley (1994), Naim (1997). 
However, one is left with a black box, the workings of which are difficult
to decipher or translate into the physical parameters of a galaxy.

A further line of approach was introduced by Abraham et al. (1994, 1996), who introduced the 
concentration index  $C$ of a galaxies image as a ratio of two radii, fixed as enclosing two 
given percentages of the total light of the system, at any given wavelength. This approximately corresponds to a
bulge to disk ratio, and was shown to correlate well with typical Hubble types. The scheme was extended by
Schade et al. (1995) to include an objective measure of the asymmetry of the projected image, through the
$A$ index, essentially the quotient of the residual of the images light after a 180 degree rotation of it
has been subtracted, to the total light. This last is particularly sensitive to the effects of
mergers and interacting systems, but not very relevant in telling apart the different classes
of generally highly 180-rotationally symmetric isolated galaxies along the Hubble sequence. The last
component of this image classification scheme is the $S$ clumpiness index introduced by 
Takamiya (1999) and Conselice (2003), a similar dimensionless index quantifying the degree of
small scale structure in an image, similar to the ratio of small scale to large scale power in a 
polynomial decomposition of the image e.g. (Kelly \& McKay 2004). This approach essentially
tackles the careful, objective, statistical description of an image through the introduction of
the dimensionless parameters expected to be the most representative of the problem. Other variants 
have appeared, for example Lotz et al. (2004) define two different statistical descriptions of the
distribution of light on the sky, and show that this allows an adequate automatic assignment of
a galaxies classical Hubble type. 

The above methods however, remain somewhat sensitive to projection and wavelength effects, but more
seriously, do not include any explicit information of the dynamics of the problem. Certainly, the 
ability to deduce a galaxies Hubble type from its $CAS$ or other careful objective, dimensionless
characterization of its image, suggests these factors vary monotonically along the Hubble
sequence. Again, it is not evident how the structural physics of the problem result in a given
point in $CAS$ space.

Monotonic trends with Hubble type suggest that after having fixed the mass, there might exist one other
parameter whose variation gives rise to the Hubble sequence. Many approaches to galaxy
formation and evolution have appeared over the years, generally reaching the conclusion that it is 
the angular momentum of a galactic system what determines its main characteristics. Some example
of which are: Sandage et al. (1970), Brosche (1973), van der Kruit (1987),
Fall \& Efstathiou (1980), Flores et al. (1993), Firmani et al. (1996), Dalcanton et al. (1997), 
van den Bosch(1998), Hernandez \& Gilmore (1998), Avila-Reese et al. (1998),  Zhang \& Wyse (2000), 
Ferguson \& Clarke (2001), Silk (2001), Bell (2002), Kregel et al. (2005). 

A second problem is what 
determines the angular momentum of a galaxy at any given time, with older modelings taking this value as 
a fundamental parameter fixed by initial conditions in the remote past, and more recent CMD 
cosmological simulations deducing this value as a consequence of the tidal fields of surrounding matter,
in combination with the formation history of a halo, with merger events and gas cooling being fundamental 
to a determination of the sometimes fluctuating value of a galactic systems angular momentum e.g.
the cosmological N-body studies of Warren et al. (1992), Cole \& Lacey (1996), Lemson \& Kauffmann (1999),
Steinmetz \& Navarro (1999) and Navarro \& Steinmetz (2002) or the analytical work of Catelan \& Theuns (1996). 
Whatever the
origin or evolution of this parameter, once the mass of a galaxy has been chosen, theoretical studies have
typically identified the angular momentum of the system as the principal determinant of a galaxies type.
We will therefore focus our attention on this parameter.

Many different models of the structure and evolution of a disk galaxy exist, with varying degrees of
detail and including diverse physical aspects of the problem e.g., asides from the above references, including
a more explicit cosmological formation scenario, Frenk et al. (1985), Frenk et al. (1988), White et al. (1991),
Mo et al. (1998),
Somerville \& Primack (1999), Maller et al. (2002) and Klypin et al (2002), to mention only but a representative few.
Here we present
the simplest possible physical treatment of the problem, not in an attempt to reproduce or understand the
details of a disk galaxy, but rather to obtain a first order approximation which might capture
the monotonic trends nicely followed by the Hubble sequence. We present a highly simplistic physical modeling
of a disk galaxy which allows to estimate its angular momentum parameter $\lambda$ (Peebles 1969) 
from readily obtainable structural characteristics.

The layout of our paper is as follows:
section 2 presents the physics of the dimensional analysis of disk galaxies, leading to an observational
estimate of the $\lambda$ parameter for any real galaxy. This parameter is then calculated for a large sample of 
galaxies in section 3, where we also explore the trends followed by different observed galactic properties 
with $\lambda$, and compare with the equivalent trends seen for the Hubble classification. In section 4
we compare our estimated $\lambda$ parameter for several simulated galaxies from various authors, and
present a discussion of our results and conclusions.

\section{Theoretical framework}

As discussed in the introduction, the good monotonic trends followed by a variety of galactic properties
along the Hubble sequence reflect not only the usefulness of it, but also suggest that to first order,
a disk galaxy can be described by only two parameters, total mass, and some other physical parameter
varying monotonically along the Hubble sequence. As mentioned in the introduction, extensive 
theoretical work leads to the angular momentum as a natural choice for this parameter.

Having chosen the angular momentum as our second parameter we now construct a simple 
physical model for a galaxy which allows to estimate this parameter directly from observed galactic properties.

In the interest of capturing the most essential physics
of the problem, the model we shall develop will be the simplest one could possibly propose, little more than
a dimensional analysis of the problem.

\subsection{Estimating $\lambda$ from observations}

We shall model only two galactic components, the first one a disk having a surface mass 
density $\Sigma(r)$ satisfying:

\begin{equation}
\label{Expprof}
\Sigma(r)=\Sigma_{0} e^{-r/R_{d}},
\end{equation} 

Where r is a radial coordinate and $\Sigma_{0}$ and $R_{d}$ are two constants which are allowed to vary from
galaxy to galaxy. The total disk mass is now:

\begin{equation}
M_{d}=2 \pi \Sigma_{0} R_{d}^{2}.
\end{equation} 

The second component will be a fixed dark matter halo having an isothermal $\rho(r)$ density 
profile, and responsible for establishing a rigorously flat rotation curve $V_{d}$ throughout the entire galaxy,
an approximation sometimes used in simple galactic evolution models e.g. Naab \& Ostriker (2005), such that

\begin{equation}
\label{RhoHalo}
\rho(r)={{1}\over{4 \pi G}}  \left( {{V_{d}}\over{r}} \right)^{2}, 
\end{equation}

and a halo mass profile $M(r)\propto r$.
Since the total mass of the disk is finite, we define the disk mass fraction as
$F=M_{d}/M_{H}$. Requiring a finite total halo mass, $M_{H}$, will now imply a truncation radius for the dark halo,
$R_{H}$ given by the equation:

\begin{equation}
M_{H} = \int_{0}^{R_{H}} \frac{V_{d}^{2}}{G} dr  \Rightarrow
\end{equation}

$$
R_{H}=\frac{M_{H}G}{V_{d}^{2}}.
$$

The disk mass fraction $F$ is expected to be of order $1/10$ or smaller (e.g. 
Flores et al. 1993, Hernandez \& Gilmore 1998),
hence, we shall use global parameters for the entire system indistinctly from halo
parameters, consistent with having ignored disk self gravity in eq(\ref{RhoHalo}).

It will be convenient to refer not to the total angular momentum $L$, but to the 
dimensionless angular momentum parameter

\begin{equation}
\label{Lamdef}
\lambda = \frac{L \mid E \mid^{1/2}}{G M^{5/2}}
\end{equation}

where $E$ is the total energy of the configuration and $G$ is Newton's
gravitational constant. $\lambda $ in fact gives the ratio of the actual angular
momentum of the system to it's maximum possible break-up value, $\lambda=1$ is hence
an upper limit.

We must now express $\lambda$ in terms of structural galactic parameters readily
accessible to observations. First we assume that the total potential energy of the galaxy
is dominated by that of the halo, and that this is a virialized gravitational structure,
which allows to replace $E$ in the above equation for $W/2$, one half the gravitational
potential energy of the halo, given by:

\begin{equation}
W=-4 \pi G \int_{0}^{R_{H}} \rho(r) M(r) r dr 
\end{equation}

use of equation (\ref{RhoHalo}) yields:

\begin{equation}
\label{Epot}
W=-V_{d}^{2}M_{H}.
\end{equation}

Assuming that the specific angular momenta of disk and halo are equal, $l_{d} =l_{H}$,
as would be the case for an initially well mixed protogalactic state, and generally 
assumed in galactic formation models e.g. Fall \& Efstathiou (1980), Mo et al. (1998).

The specific angular momentum of the disk, following
the assumption of a rigorously flat rotation curve and eq(\ref{Expprof}), will now be
$l_{d}=2 V_{d} R_{d}$. We can now replace $L$ for $M_{H} l_{d}$. Introducing this last result
together with eq(\ref{Epot}) into eq(\ref{Lamdef}) yields:

\begin{equation}
\label{Lamhalo}
\lambda=\frac{2^{1/2} V_{d}^{2} R_{d}}{G M_{H}}.
\end{equation}

Finally, we can replace $M_{H}$ for $M_{d}/F$, and introduce a disk Tully-Fisher relation:

\begin{equation}
\label{TullyF}
M_{d}=A_{TF} V_{d}^{3.5}
\end{equation}

into eq(\ref{Lamhalo}) to yield:

\begin{equation}
\label{LamTully}
\lambda= \left( \frac {2^{1/2} F}{A_{TF}} \right) \frac{R_{d}}{G V_{d}^{3/2}}.
\end{equation}

The existence of a general baryonic Tully-Fisher between the total baryonic component and $V_{d}$ 
of the type used here, rather than a specific Tully-Fisher relation involving total magnitude in any particular band,
is supported by recent studies, in general in agreement with the 3.5 exponent we assume (e.g. Gurovich et al. 2004
or Kregel et al. 2005 who find 3.33 $\pm$ 0.37).
All that remains is to chose numeric values for $F$ and $A_{TF}$, to obtain an estimate of the
spin parameter of an observed galaxy in terms of structural parameters readily accessible to
observations, $R_{d}$ and $V_{d}$.

Taking the Milky Way as a representative example, we can use a total baryonic mass of
$1 \times 10^{11} M_{\odot}$ and $M_{H} = 2.5 \times 10^{12} M_{\odot}$ (where the estimate of the
baryonic mass includes all such components, disk, bulge, stellar halo etc., e.g. Wilkinson \& Evans 1999, 
Hernandez et al. 2001) as
suitable estimates to obtain $F=1/25$. For a rotation velocity
of $V_{d}=220 km/s$, the above numbers imply $A_{TF}=633 M_{\odot} (km/s)^{-3.5}$, this in turn
allows to express eq(\ref{LamTully}) as:

\begin{equation}
\label{LamObs}
\lambda=21.8 \frac{R_{d}/kpc}{(V_{d}/km s^{-1})^{3/2}},
\end{equation}

which is the final result of this sub-section. The above equation allows a direct estimate of
$\lambda$, a dimensionless numerical parameter with a clear physical interpretation for any observed galaxy, 
with no degree of subjectivity and little sensitivity to orientation effects. The only ambiguity which remains
is the definition of the wavelength at which $R_{d}$ is measured. Recently de Grijs (1998) and 
Yoachim \& Dalcanton (2005) and have measured
significant variations in $R_{d}$ with wavelength. In the spirit of obtaining global mass distribution
parameters, it is clear that observations in the r-band or near IR are required for the measure of
$R_{d}$ necessary in eq(\ref{LamObs}). Also, given the Tully-Fisher relation, it is clear that
$V_{d}$ in eq(\ref{LamObs}) can be replaced for total magnitude on a given band, for cases where
the rotation velocity is not available. It is clear from its derivation that equation (\ref{LamObs})
is only an approximate way of inferring the $\lambda$ parameter of an observed galaxy, intended as 
an objective and quantitative improvement on current classification schemes, and not as a detailed
measurement of the angular momentum of a real galaxy. If one intended a more careful measurement of 
$\lambda$, a full dynamical modeling of the rotation curve and bulge and disk matter distribution
would be required, of the type found in e.g. Tonini et al. (2005).

For the Galactic values used above, equation (\ref{LamObs})
yields $\lambda_{MW}=0.0234$, in excelent agreement with recent estimates of this parameter, e.g.
through a detailed modeling of the Milky Way within a CDM framework Hernandez et al. (2001) find
$\lambda_{MW}=0.02$.

\subsection{Expected scalings with $\lambda$}

We now study the expected scaling between $\lambda$ and other observables of a disk galaxy, 
for example colour, gas fraction, star formation activity, degree of flatness of the disk and
bulge to disk ratios. The main ingredient will be the Toomre parameter,

\begin{equation}
\label{Toomre}
Q(r)=\frac{\kappa(r) \sigma(r)}{\pi G \Sigma(r)},
\end{equation}

where $\kappa(r)$ is the epicyclic frequency at a given radius, and
$\sigma(r)$ is the velocity dispersion and $\Sigma(r)$
the disk surface density of any given galactic disk component.
This parameter measures the degree of internal gravitational stability of the disk,
with $\kappa(r)$ accounting for the disruptive shears induced by the differential rotation,
$\sigma(r)$ modeling the thermal pressure of a component, both in competition with $\Sigma(r) G$, the self gravity
of the disk. Values of $Q \leq 1$ are interpreted as leading to gravitational instability in a 'cold'
disk. In the interest of simplicity of the description, and bearing in mind that the angular 
velocity and the epicyclic frequency are within a factor of order unity of
each other, e.g. in a flat rotation curve, $\kappa(r) = \sqrt{2} \Omega(r)$, 
we shall replace $\kappa(r)$ in eq(\ref{Toomre}) with $\Omega(r) = V_{d}/r$, the angular velocity of the disk.        

The star formation processes in the disk have often been thought of as forming a self regulated cycle. 
The gas in regions where $Q<1$ is gravitationally unstable, this leads to collapse and the triggering of
star formation processes, which in turn result in significantly energetic processes. The above include
radiative heating, the propagation of ionization fronts, shock waves and in general an efficient
turbulent heating of the gas media, raising $\sigma$ locally to values resulting in $Q>1$. This
restores the gravitational stability of the disk and shuts off the star formation processes. On timescales
longer than the few $\times 10$ million years of massive stellar lifetimes, an equilibrium 
is expected where star formation proceeds at a rate equal to that of gas turbulent dissipation, at
time averaged values of $Q\sim 1$. Examples of the above can be found in Dopita \& Ryder (1994), 
Koeppen et al. (1995), Silk (2001).

For example, Silk (2001) uses ideas of this type to derive the local 
volume density of star formation $\dot{\rho}_{sf}$ as: 

\begin{equation}
\label{SFSilk}
\dot{\rho}_{sf} = \Omega \rho_{g} \sigma_{g},
\end{equation}

Where volume densities will be given by $\rho=\Sigma/h$, $h$ the disk scale height.
Employing the Toomre criterion at $Q=1$ to rewrite the gas velocity dispersion in terms of the gas surface
density, equation (\ref{SFSilk}) may be written as follows:

\begin{equation}
\label{Shmidt}
\dot{\Sigma}= \pi G \Sigma^2_{g}
\end{equation}

The above represents a simple derivation of a Schmidt law, other similar options have been proposed 
that lead also to Schmidt laws with exponents between 1-2, e.g. Firmani et al. (1996). 
We now use a characteristic time for the star formation process in the 
disk $\tau_{sf} = \Sigma_{g} / \dot{\Sigma}$, that gives an idea of the time a galaxy
can support a star formation rate of constant magnitude; if its value is small, the duration of star formation processes 
will be short, the population presently old and consequently, the galaxy will be red and have a low gas fraction. 
On the other hand, a galaxy with large $\tau_{sf}$ will experience a long star formation duration, and show a
young population resulting in bluer colours and a high gas fraction.
 
From the above relation $\tau_{sf}$ becomes 

\begin{equation}
\label{tauSF}
\tau_{sf} = \frac{\pi G}{\Sigma_{g}}=\frac{2 \pi^2 G R^2_{d}}{M_{d}}
\end {equation}
 
Notice that at fixed disk mass (or fixed $V_{d}$, through the Tully-Fisher relation), from eq(\ref{LamObs})
$\lambda$ will scale with $R_{d}$, which translates into a relation between $\tau_{sf}$ and
$\lambda$:

\begin {equation}
\label{TauSF}
\tau_{sf} \propto \lambda^2.
\end {equation}

In the above, representative values of $\kappa(r)$, $\Omega(r)$, $\Sigma(r)$, $\sigma(r)$ and $\rho_{g}(r)$
 have been assumed, for
example, evaluating all variables at $r=R_{d}$. If one assumes a different star formation scheme from what was used above,
equation (\ref{TauSF}) will change. However, for a variety of similar simple schemes, the proportionality given will 
be altered only in the change of the exponent, which will remain $>1$. Alternatively, one could start from an
empirical Schmidt law of power $n$, a scaling $\tau_{sf} \propto \lambda^n$ will always result, which does not
alter our conclusions provided $n>1$, a general trait of Schmidt laws found in the literature e.g. Silk (2001).

Large values of $\lambda$ will correspond to long star formation periods; for this case we expect to have galaxies 
with young populations and looking relatively blue. Galaxies with low $\lambda$ will have short star
formation periods; for them we will hardly see young stars and they will look red and gas poor. A scatter
on this trend will be introduced by variations in total mass, together with a systematic reduction in
$\tau_{sf}$ in going to larger masses (c.f. equation \ref{tauSF}), in accordance with larger average
galactic masses found at earlier types.

The ratio of disk scale height to disk scale length, $h/R_{d}$, is another measurable characteristic 
of galaxies which it is easy to show, will also scale with $\lambda$. Assuming a thin disk, virial equilibrium 
in the vertical direction (Binney and Tremaine 1994) yields a relation between $h$ and $\Sigma$,

\begin{equation}
\label{VertEq} 
h= \frac {\sigma_{g}^{2}} { 2 \pi G \Sigma}. 
\end{equation}

We use this relation for 
$h$ to replace the gas velocity dispersion appearing in equation (\ref{Toomre}) for a combination of 
$h$ and the surface density. The dependence on $\Sigma$ is replaced by a dependence on $M_{d}$, 
$M_{H}$ and $\lambda$ to get a new expression for the Toomre's stability criterion, which evaluating radial
dependences at $r=R_{d}$ yields,

\begin {equation}
Q^2= e 2^{5/2} \left( \frac{M}{M_{d}} \right) \left( \frac{h}{R_{d}} \right) \lambda
\end {equation}

With $F=1/25$, evaluating at $Q=1$, the stability threshold suggested by self-regulated star formation cycles, 
the ratio $h/R_{d}$ is obtained as:

\begin {equation}
\label{hRratio}
\frac{h}{R_{d}} =  \frac{1}{390 \lambda},
\end {equation}

a simplified version of the result of van der Kruit (1987).
For the Galactic value derived above of $\lambda_{MW}=0.0234$, equation (\ref{hRratio}) gives
$R_{d}=9 h$, not in conflict with parameters for the Milky Way.
For galaxies with large values of $\lambda$ we expect thin systems, while galaxies with small values of $\lambda$ 
will show thick disks, as the observed trend of ($h/R_{d}$) decreasing in going from early-type disks
to late type galaxies e.g. de Grijs (1998), Yoachim \& Dalcanton (2005).

\begin{figure*}
\epsfig{file=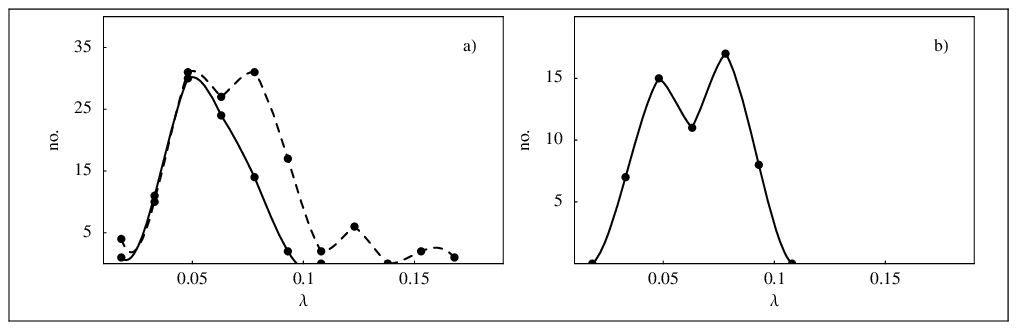,angle=0,width=18.1cm,height=6.0cm}
\@ \textbf{Figure 2.}\hspace{5pt}{
\begin{flushleft}\textbf{a)} The solid line represents the distribution of values of $\lambda$ through eq(\ref{LamObs})
for Sb galaxies in the ASSSG sample, the dashed line gives the equivalent distribution for Sc galaxies.
\textbf{b)} Distribution of values of $\lambda$ for the Sbc galaxies in that same sample. Dots showing the binning.
\end{flushleft}}
\end{figure*} 

Another important observable characteristic of the morphology of a galaxy, which shows
clear trends with Hubble type, is the bulge to disk ratio $B/D$.
From observations it is clear that the central regions of the disk of a galaxy are dynamically ``hot'', because of the
large velocity dispersion present in comparison to the tangential rotation velocity. With this in mind, 
we introduce a kinematic definition of the bulge (e.g. Hernandez 2000) as the region where the velocity 
dispersion is comparable to the circular velocity of the disk. The ratio will be given by

\begin{equation}
\frac{B}{D}=\frac{\int_{0}^{a} 2 \pi \Re \Sigma(\Re) d\Re}{\int_{a}^{\infty} 2 \pi \Re \Sigma(\Re) d\Re},
\end{equation}

with $\Re$ defined as $\Re=r/h$ and $a$ a dimensionless parameter $\geq 1$. The bulge mass is the mass 
contained from the center of the disk to a distance equal to a multiple of $h$; $ah$, and the disk mass 
is the remainder. The dependence of the $h/R_{d}$ ratio is changed for a dependence in $\lambda$, using 
equation (\ref{hRratio}). The above equation leads to:

\begin{equation}
\label{BtD}
\frac{B}{D} = \frac{L(e^{a/L} -1)-a}{L+a}
\end{equation}

where we have introduced $L=390 \lambda$. We expect to find galaxies with large bulges 
for low $\lambda$ values. Of course, other processes 
which we have not considered, such as all manner of galactic interactions, will result in angular momentum transfer
and generally the induction of matter flows towards the bottom of the potential well, i.e. an increase in the
$B/D$ ratio. This leads us to expect significant spread.

Also, the analysis of this section implies an increase in values of $R_{d}$ and a decrease
in $\Sigma_{0}$ for more flattened, high $\lambda$ disks. This could account for the
observation of a negative correlation between  $R_{d}$ and $\Sigma_{0}$ found by
Kregel et al. (2005). These authors also find through dynamical modeling of galaxies having
2D photometry, that the mass to luminosity ratios of galaxies increase for more flattened
(low $h/R_{d}$ ratios) disks. This last trend appears naturally if we think of flattened disks
as coming from high $\lambda$ values, leading to large $R_{d}$ and hence galactic disks that
extend far out into their dark halos, out to regions where the dominance of the dark
component is increasingly obvious. Compounding the above is the fact that volume density of 
disks scales with $(h R_{d}^{2})^{-1}$, hence since $h$ scales with $\lambda^{-1}$ and $R_{d}$
does so with $\lambda ^{2}$,
to this level of the analysis, the volume density of the disk will decrease linearly with $\lambda$,
leading to a decrease in the gravitational potential of the disk. Variations in mass will again
introduce scatter into the above trends.

As we have shown in this sub section, in general, galaxies with small values of $\lambda$ 
would be expected to be red, have a low gas content, show little present star formation, thick disks 
and typically large bulge to disk ratios, the defining traits of early spirals.
On the other hand, for large values of $\lambda$, we expect blue, thin galaxies
showing the small bulge to disk ratios of late-type and low surface brightness (LSB) disks.
It appears reasonable to expect the observational estimates of $\lambda$ of eq(\ref{LamObs}) to show
all the trends of the Hubble sequence, indeed, from the point of view of galactic formation scenarios,
it is $\lambda$ what defines a galaxy's type (e.g. Fall \& Efstathiou 1980 van der Kruit 1987, 
Flores et al. 1993, Dalcanton et al. 1997, Hernandez \& Gilmore 1998 and van den Bosch 1998). 

\begin{figure}
\epsfig{file=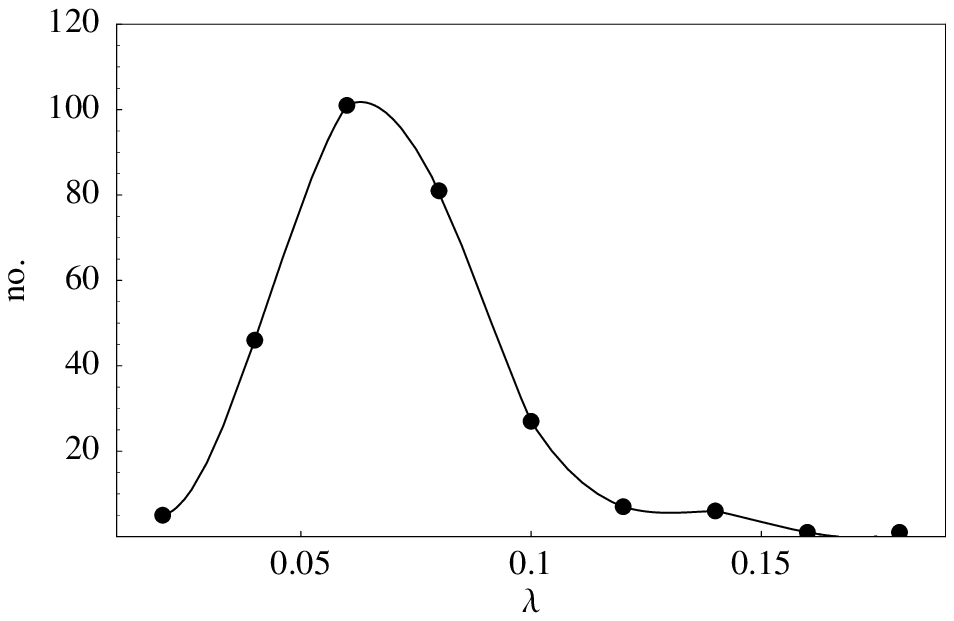,angle=0,width=8.5cm}
\@ \textbf{Figure 1.}\hspace{5pt}{Distribution of values of $\lambda$ through eq(\ref{LamObs}) for the
complete ASSSG sample, with dots showing the binning}
\end{figure}

Notice that this simple model will be valid independently
of the formation scenario of a disk galaxy. Whether low values of $\lambda$ for early spirals 
are fixed in the remote past as initial conditions, or the result of cancellation of angular
momentum due to repeated mergers, equation(\ref{LamObs}) will still give a reasonable estimate of the value of
$\lambda$ for a disk galaxy at a given time of observation. The model presented in this section 
isolates the main physical ingredients responsible for the correlations between the galactic observables
described. Other more complex and complete models have been presented in the past, which however do not
deviate significantly from the basic assumptions of section (2.1), and consequently reach conclusions not
drastically different from the ones of section (2.2), if we limit the comparison to only the specific
aspects of the problem being treated here. Examples of the above are the analytical schemes of
van der Kruit (1987), refined later by Dalcanton et al. (1997), or the more numerical schemes of 
Fall \& Efstathiou (1980), Flores et al. (1993), before the arrival of fuller hydrodynamical and
cosmological simulations.

\section{Observational data}

To compare the relations obtained in the preceding section with observational data of real galaxies
we used two samples of galaxies; the Atlas for Structural Studies
of Spiral Galaxies (ASSSG) of the Spiral Galaxy data base of Courteau (1996, 1997) and a sample taken from 
de Grijs (1998) based on the Surface Photometry Catalog of the ESO-Uppsala Galaxies (ESO-LV; 
Lauberts \& Valentijn 1989), and the subsequent re-analysis of these data by Kregel et al. (2002),
henceforth dGKK. This last work considers carefully issues related to dust contamination, 
projection and radial variations in disk scale height.
The ASSSG consists of a data base of 304 late-type spiral galaxies including structural parameters such as
rotation velocities and $R_{d}$, measured in the r-band. This last point is important, as we require
a measure of the extent of the disk mass, and red bands are more sensitive to underlying old stellar populations
and are little affected by dust.
 
From this sample we used the reported disk circular velocity measured at 
2.15 disk scale-lengths, the corrected disk scale length,
the color term B-R and the ratio of fitted disk light to total measured light for the whole galaxy. 
This last to estimate $B/D$ ratios. The spectral 
band used for these measurements was the R band. All the galaxies which show some kind of interaction were suppressed 
for the analysis, together with spherical and irregular galaxies. One difficulty of using this sample, 
in connection with our present goals, is the lack of early type spirals. This will make trends with Hubble 
type appear harder to detect, as it is often between the Sa and the latter types that the largest changes occur.
However, the uniform treatment of the data, the large sample and the long observational wavelengths make 
the ASSSG sample worthwhile. If some trends can be seen even in the absence of Sa galaxies, as it is
indeed the case, it is to be expected that their inclusion would only strengthen further our conclusions.

The dGKK sample consists of 33 edge-on 
galaxies from which we used the I-band exponential scale height and exponential scale length and the maximum 
rotational velocity. Again, it is the red wavelength of observation that makes this particular 
sample relevant to our analysis.

As mentioned in the preceding section, we expect a correlation between the $\lambda$ parameter and the 
Hubble type, since the analysis of section 2 leads one to suspect that it is this what 
primarily determines the properties of the galaxy
which are taken into account for its classification. In figure 1 we present the distribution of the 
ASSSG sample as a function of $\lambda$, estimated through eq(\ref{LamObs}). 
For the entire sample, the mean value is $< \lambda >=0.0645$. 
We see an approximately Gaussian shape, with a considerable extension towards the larger values of
$\lambda$. We do not explore the details of this distribution further, as the ASSSG sample is not
intended to be complete or statistically unbiassed in any sense. However, use of eq(\ref{LamObs})
in one such sample should prove useful in obtaining an empirical distribution of $\lambda$ parameters,
to compare for example, against predictions from cosmological models. These models yield theoretical
distributions for galactic values of $\lambda$ resulting from tidal interactions of the material 
acreting onto a protogalaxy and global surrounding dark matter tidal fields,
which are sensitive to cosmological parameters such as
$\Omega_{\Lambda}$ and the details of the initial fluctuation spectrum (e.g. Warren et al. 1992, Cole \& Lacey 1996, 
Catelan \& Theuns 1996 and Lemson \& Kauffmann 1999). 
The above parameters could in principle then be constrained through
empirical inferences of the present day distribution of $\lambda$ inferred through equation (\ref{LamObs})
and a large unbiased sample of galaxies.

Second, we divide the ASSSG sample by Hubble type into three groups; Sb, Sbc and Sc. The result is shown in 
figure 2, where the continuous curve in panel a) shows the distribution of values of $\lambda$ for the Sb's and the
dashed curve for the Sc's. We see that as expected, Sc's are characterized by larger values of
$\lambda$, although significant overlap occurs. Panel b) gives the $\lambda$ distribution for
the subclass Sbc's, by comparing to panel a) it is interesting that this class appears as
a superposition of the previous two classes, with the exception of the tail towards large
$\lambda$s seen in the Sc sample. This last point highlights the difficulty in making
detailed distinctions among sub classes in the Hubble sequence, and agrees with the conclusions of
Ellis et al. (2005), who analyze and classify automatically through a variety of elaborate statistical
analysis on many galactic parameters a large sample of galaxies, and conclude that only 
two definitive classes are naturally suggested by the data, late and early spirals.

In general, the bulk of a galactic population moves to larger values of $\lambda$ for later types,
the mean value for each group is: $<\lambda>_{Sb}=0.0574, <\lambda>_{Sbc}=0.0649, <\lambda>_{Sc}=0.0689$.

\begin{figure}
\epsfig{file=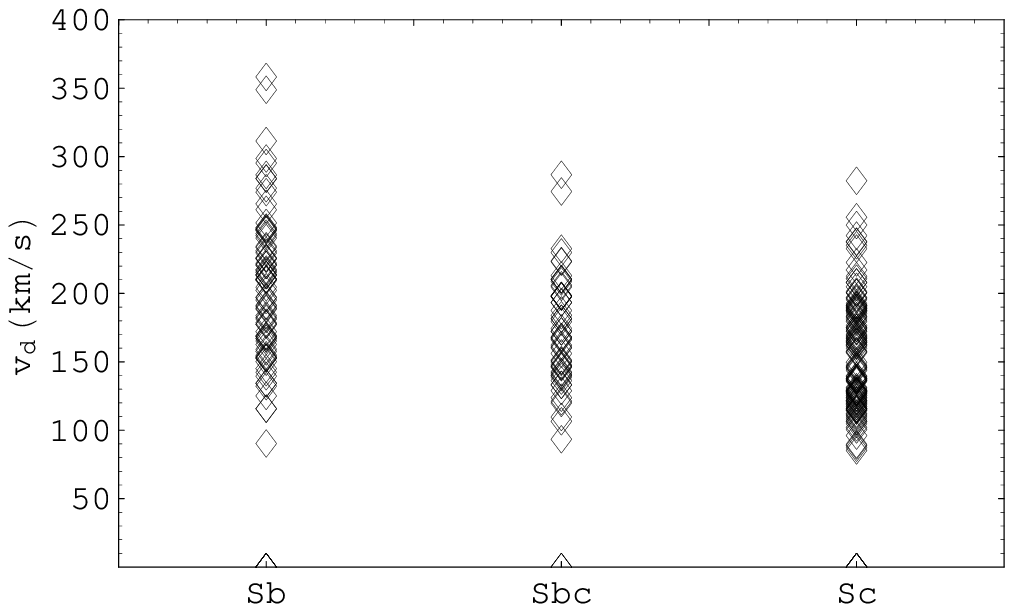,angle=0,width=8.5cm}
\@ \textbf{Figure 3.}\hspace{5pt}{Hubble type as a function of the circular velocity for the galaxies in the
ASSSG sample.}
\end{figure}

Next we analyze and compare the statistical distributions of $V_{d}$ as functions of Hubble type and
$\lambda$, also for the ASSSG. In figure 3 we present the
distribution of galactic rotation velocities $V_{d}$ with morphological type.
Earlier types present a wide dispersion of velocities, which diminishes slightly for the later types,
where lower mean values are found. For this analysis we obtained mean values for every sub-sample; 
$<v>_{Sb}=205.41  km/s$, $<v>_{Sbc}=172.42  km/s$ and $<v>_{Sc}=155.95  km/s$, a monotonic trend is clear,
although very significant overlap is present. This result is well known, and reflect the similar trend
of decreasing total luminosities with Type, e.g. Roberts \& Haynes (1994), Ellis et al. (2005).

Figure 4 gives a plot of $V_{d}$ vs. $\lambda$ for the galaxies appearing in figure 3. We see that
despite considerable dispersion and overlap, galaxies with large values of $\lambda$ are restricted
to the lower values of $V_{d}$, while at low values of $\lambda$ a much wider range of $V_{d}$ is
present. The mean $V_{d}$ at each $\lambda$ decreases with $\lambda$, as is seen when plotting against 
Hubble type in figure 3.

An important measurable parameter for the classification of a galaxy is its color, giving
information on the evolutionary state of the stellar population. One of the clearest
trends seen along the Hubble sequence is precisely the shift to bluer colours when going towards the later types
(e.g. Roberts \& Haynes 1994, Park \& Choi 2005). The ASSSG sample shows this expected relation 
clearly in figure 5, where we plot B-R colour for the different types.

\begin{figure}
\epsfig{file=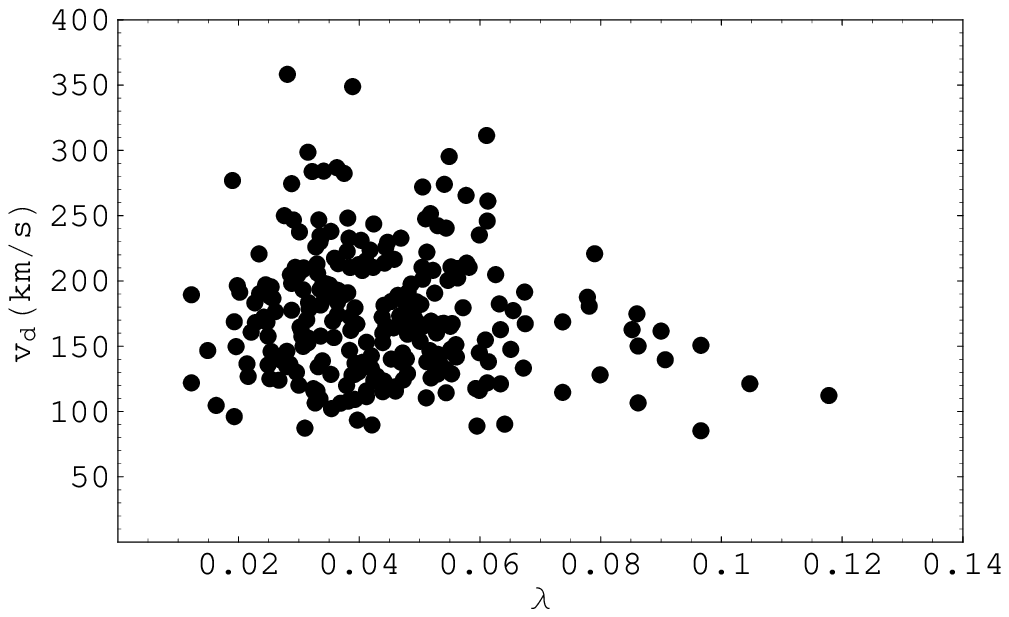,angle=0,width=8.5cm}
\@ \textbf{Figure 4.}\hspace{5pt}{Values of the circular velocity plotted against $\lambda$ through eq(\ref{LamObs})
for the galaxies in the ASSSG sample.}
\end{figure}

\begin{figure}
\epsfig{file=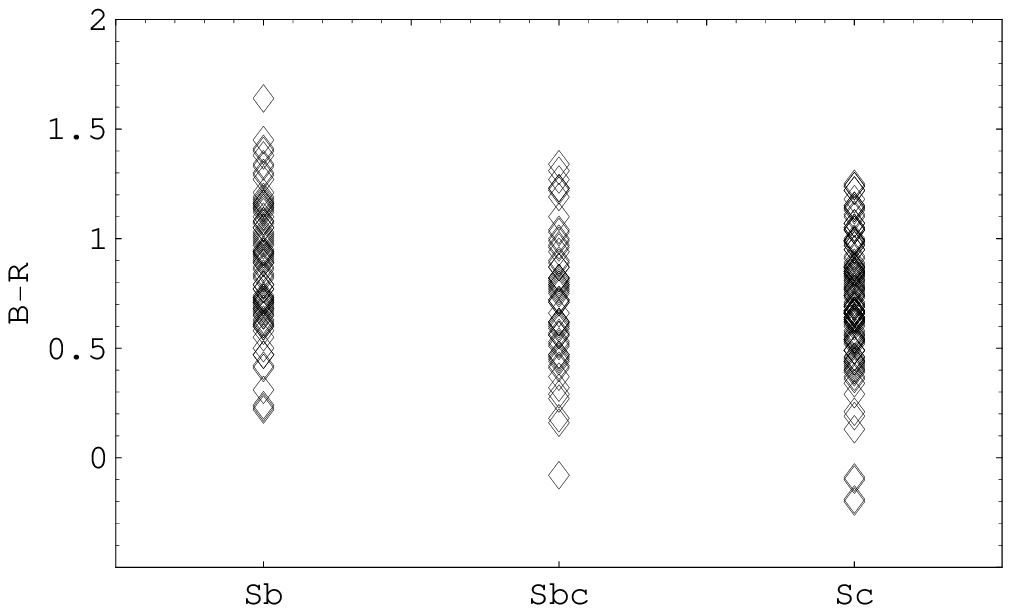,angle=0,width=8.5cm}
\@ \textbf{Figure 5.}\hspace{5pt}{Relation between the Hubble type and the B-R color term,
for the galaxies in the ASSSG sample.}
\end{figure}

In section 2, we proposed a relation between $\lambda$ and the color of a galaxy supported by the idea of 
the existence of a characteristic timescale for the star formation process, which should increase
with $\lambda$. In order to examine the validity 
of our statement we present the relation of these two parameters for the ASSSG sample in figure 6. It is clear 
that the sample follows a marked tendency, the B-R term decreases with $\lambda$,
which is in agreement with our predictions. The agreement of this trend with the equivalent one seen for
Hubble types reinforces the idea that the single most important galactic parameter which is varying along the Hubble 
sequence is precisely $\lambda$.

The analysis of the distribution of $h/R_{d}$ ratios was made with the dGKK sample. 
It contains galaxies with 
more morphological types; from Sa to Scd galaxies. This  spread of types allow us to see clearly if there is any 
kind of tendency regarding the $h/R_{d}$ ratio and the Hubble types. Figure 7 shows early galaxies 
presenting varied values 
for the $h/R_{d}$ ratio, although the means show a clear diminishing trend in going to latter types,
very similar to what was seen for $V_{d}$ and colour. Late type galaxies at the Sc end clearly show
on average much thinner disks than the earlier types, as confirmed by several authors over the last few years e.g. 
Yoachim \& Dalcanton (2005). It is interesting to note that in the
re-analysis of the de Grijs (1998) data by Kregel et al. (2002) not only do the values of $R_{d}$ and $h$
change, but perhaps surprisingly, the Hubble types of the galaxies are significantly shifted around, sometimes
by more than 1 subtype. This last point highlights again the subjective nature of ``type'' classification
schemes for galaxies. 

\begin{figure}
\epsfig{file=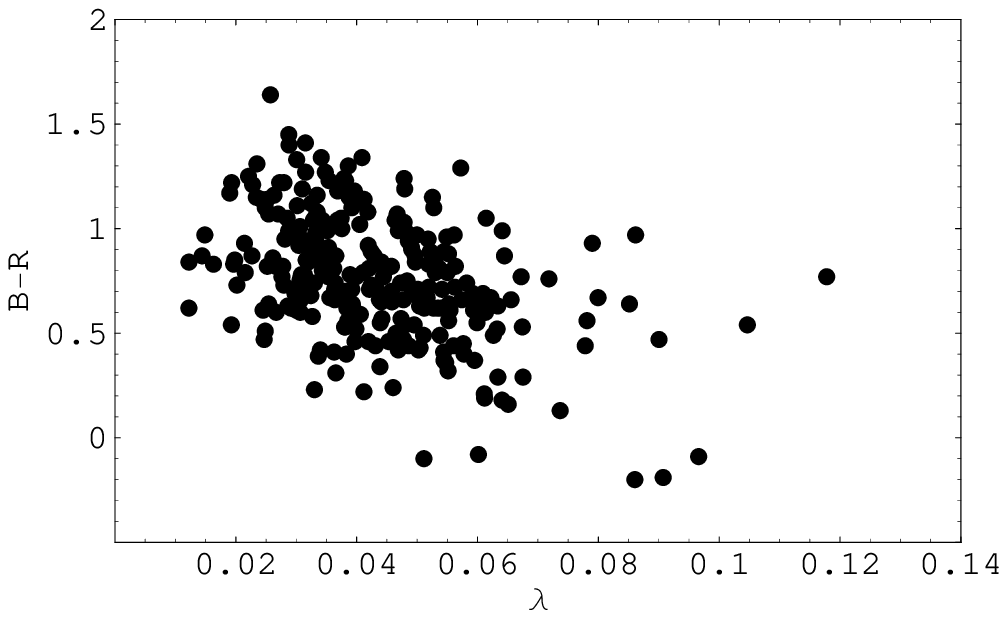,angle=0,width=8.5cm}
\@ \textbf{Figure 6.}\hspace{5pt}{Values of the B-R color term plotted against $\lambda$ through  eq(\ref{LamObs})
for all the galaxies in the ASSSG sample.}
\end{figure}

\begin{figure}
\epsfig{file=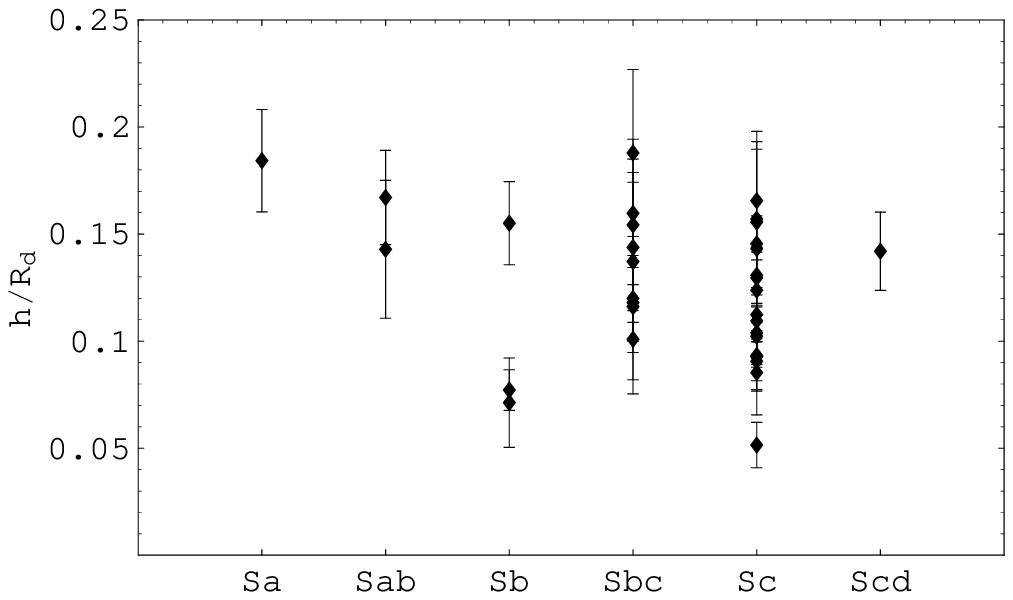,angle=0,width=8.5cm}
\@ \textbf{Figure 7.}\hspace{5pt}{$h/R_{d}$ ratios for galaxies according to their Hubble type,
for galaxies in the dGKK sample.}
\end{figure}

The complementary plot to figure 7 is given in figure 8, where instead of presenting 
$h/R_{d}$ vs $\lambda$, we use a logarithmic plot to better detect the presence of a relation
between these two parameters of the type the dimensional analysis of section 2 leads us to expect, 
equation(\ref{hRratio}). The line drawn in figure 8 is the best fit straight line for the sample,
having a slope of $-0.4 \pm 0.9$, not at odds with value of $-1$ predicted by the simple model. 
The expectations of the model are seen to be generally borne out by the data, albeit the large
errors present.

The last galactic property which we analyze is the bulge to disk ratio $B/D$. For this we return to the 
ASSSG sample, and attempt an estimate of this parameter from the reported total light, and the reported
extrapolation for a total disk light coming from the iterpolated central disk brightness and 
measured disk scale length values,
by assuming that the total measured light is the sum of the disk light and the light coming 
from the bulge. This is of course not ideal, as for example, truncated disks will give artificially
large values of the $B/D$ ratio.  This and other effect however, will affect  
all the galaxies of the sample in the same way. Therefore, the error is of the same order of magnitude for each 
and it allows us to look for and compare trends against both Hubble type and $\lambda$, although
the actual values of $B/D$ ratios might be off, and the dispersion artificially broadened.

Figure 9 gives these $B/D$ ratios against Hubble type, the extent of values found for each class is comparable, but 
the means show a very clear trend, with Sc' clustering at low values of $B/D$. This trend is again well
known, and in fact, small bulges are one of the defining characteristics of late type galaxies.

\begin{figure}
\epsfig{file=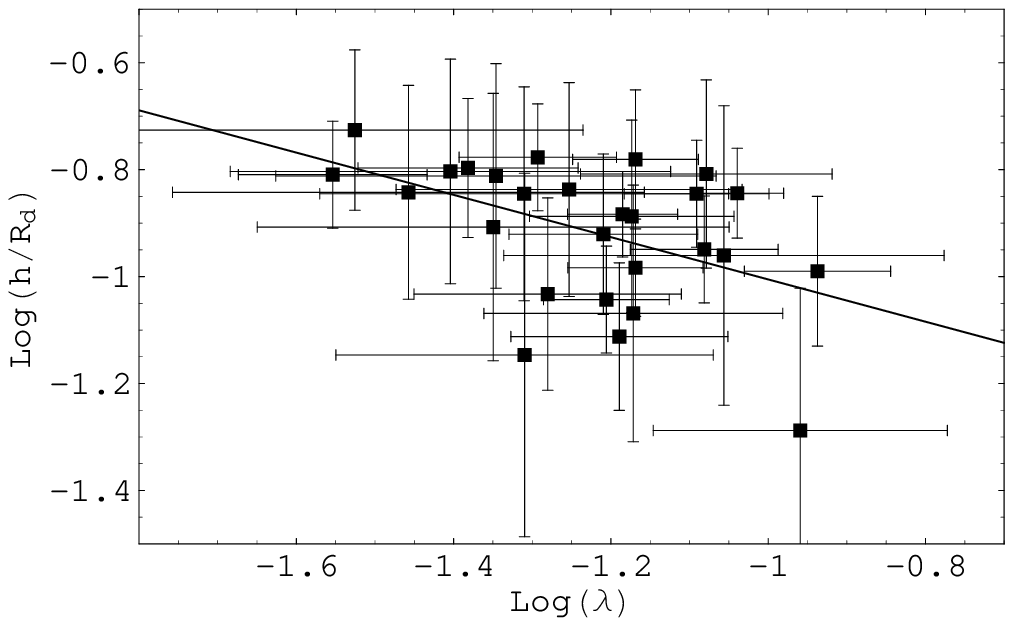,angle=0,width=8.5cm}
\@ \textbf{Figure 8.}\hspace{5pt}{Logarithmic plot of the relation between $\lambda$ and $h/R_{d}$,
for galaxies in the dGKK sample, the solid line shows a linear fit to the data, yielding
a slope of $-0.4 \pm 0.9$. }
\end{figure}

\begin{figure}
\epsfig{file=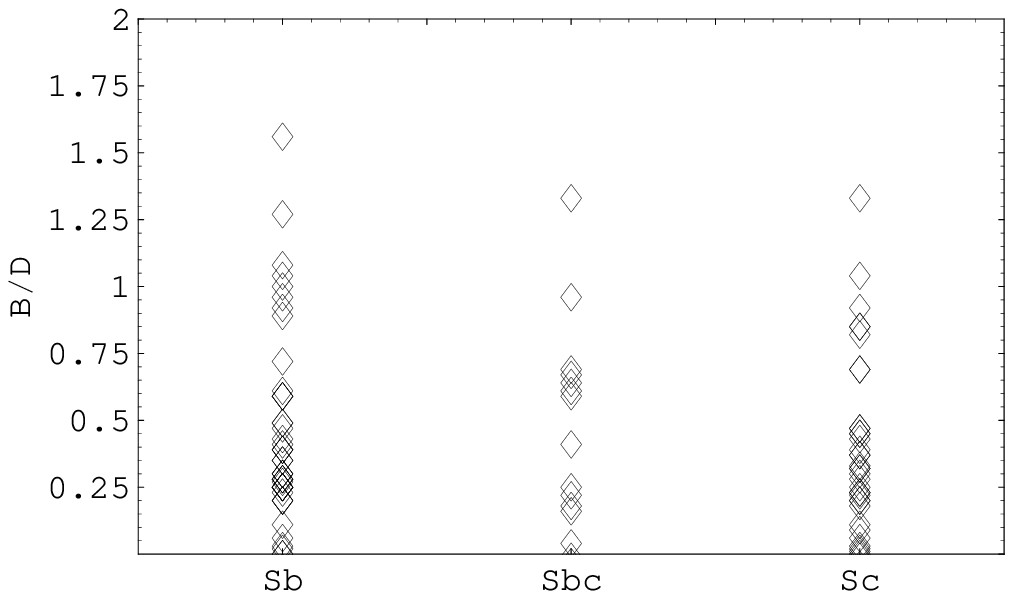,angle=0,width=8.5cm}
\@ \textbf{Figure 9.}\hspace{5pt}{$B/D$ ratios for galaxies classified by their Hubble type, for the ASSSG
sample.}
\end{figure}

The final plot of this section, figure 10, gives the same values of the $B/D$ ratio of figure 9,
but plotted this time against the value of $\lambda$ derived for each galaxy in the ASSSG
sample through eq(\ref{LamObs}). We again see something very similar to what the corresponding
'type' figure shows. Indeed, all the galactic properties examined show very similar trends, large dispersions
which drop towards later types, with means that show distinctive and very similar decreasing trends with either 
type or $\lambda$. The close equivalence of trends against Hubble type and $\lambda$ strongly supports the idea
that a combination of spatial extent (through $R_{d}$) and dynamics (through $V_{d}$) given by the dimensionless
$\lambda$ parameter of eq(\ref{LamObs}) most closely yields the monotonically changing physical parameter
of a galaxy responsible for its Hubble type. All the trends seen against $\lambda$ are noisy, but not more so
than the corresponding ones against Hubble type, and often better, with the advantages of $\lambda$ being objective, 
readily obtainable through the simple relation of eq(\ref{LamObs}) and representing a dimensionless parameter
having a direct physical interpretation.

\section{Discussion}

In this final section we discuss some aspects of the ideas proposed, and include a small comparison
of equation(\ref{LamObs}) against CDM galactic computer formation codes. A physical model simplified to
the extreme, as what was presented in section 2, will only be of any use if the few ingredients which remain
manage to capture the fundamental physical processes responsible for the aspects of the problem one is
interested in. For example, in a strong explosion, the first phases are completely determined by the energy
released and the density of the medium where the explosion takes place, these two numbers and a trivial dimensional
analysis suffice to obtain the Sedov-Taylor solution, which is an accurate description of the problem. Indeed,
complex hydrodynamical computer codes tracing the energetics and dynamics of shocks and explosions are often 
tested and calibrated in their myriad numerical details against such a simplistic analytic solution.

\begin{figure}
\epsfig{file=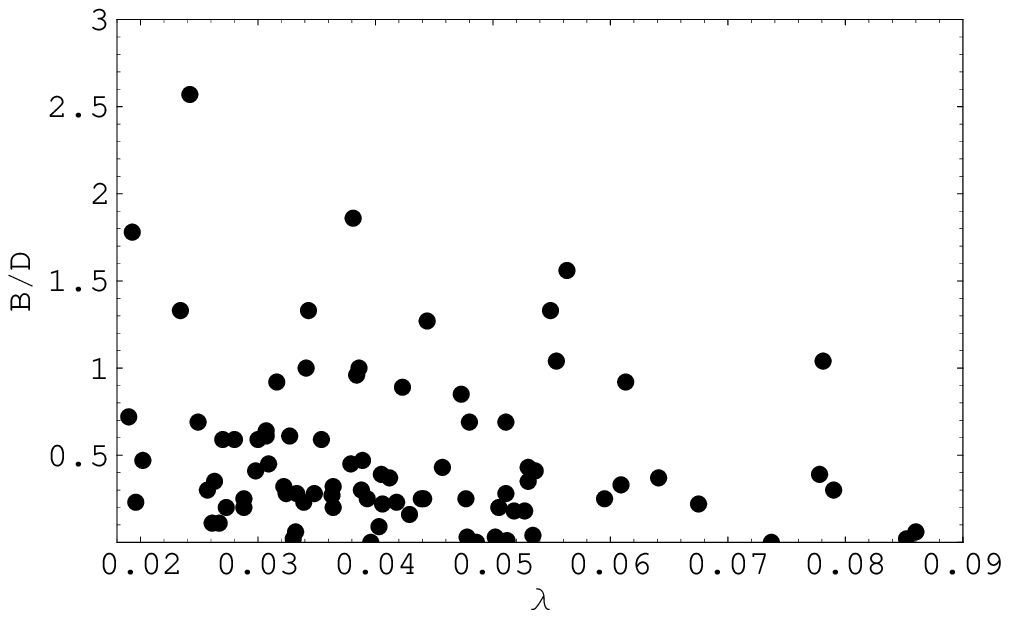,angle=0,width=8.5cm}
\@ \textbf{Figure 10.}\hspace{5pt}{PLot of $B/D$ ratios for galaxies in the ASSSG sample, as a function of 
their inferred values of $\lambda$ from eq(\ref{LamObs}).}
\end{figure}

We must now check our estimate of $\lambda$ in cases where this parameter is in fact known, galaxy formation
simulations where $\lambda$ acts as an input parameter. We use results from two groups, the Firmani et. al (1998)
approach, models calculated in connection with a study of star formation in disk galaxies applied to the 
Milky Way (Hernandez et al. 2001), and the models published by van den Bosch (2000). A very brief description
of such models is included below.

The first of these models
includes a very wide range of physics, initial conditions are supplied by an statistical sampling of
a cosmological primordial fluctuation spectrum, giving a mass aggregation history
of gas and dark matter. No Tully-Fisher type of relation is assumed {\it a priori}, indeed, this codes
attempt to recover such a relation as a final result of the physics included.
A fixed value of $\lambda$ is assumed for the infalling material which settles into a disk
in centrifugal equilibrium and differential rotation, viscous friction results in radial flows of 
matter and angular momentum. The redistribution of mass affects the rotation curve in a self-consistent way, through
a Poisson equation including the disk self gravity and a dark halo which responds to mass redistributions through
an adiabatic contraction. The star formation is followed in detail, with an energy balance cycle
of the type discussed in section 2 being introduced. 

\begin{figure}
\epsfig{file=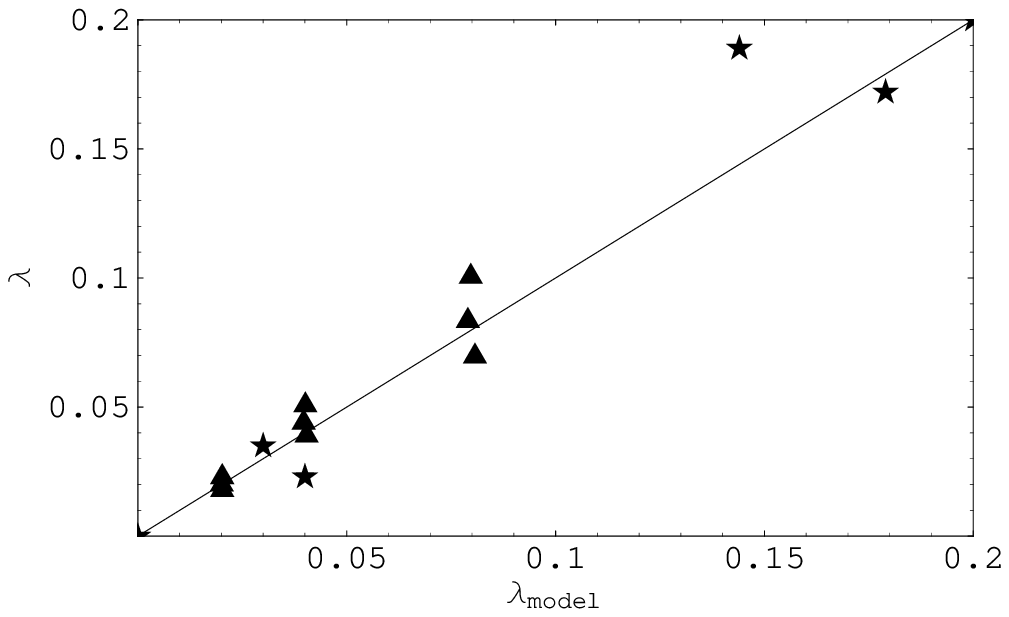,angle=0,width=8.5cm}
\@ \textbf{Figure 11.}\hspace{5pt}{A comparison of the input value of $\lambda$ in a series of cosmological
CDM galactic evolution scenarios, $\lambda_{model}$, and $\lambda$ calculated on the final result of the
given simulations through eq(\ref{LamObs}). The triangles are models from Hernandez et al. (2001) and stars from 
van den Bosh (2000).}
\end{figure}

A simple population synthesis code then traces the luminosities and colours
of the various stellar populations at all radii and times. The details of the van den Bosch (2000) models
vary in numerical approaches, resolution, time step issues and the level of approximation and inclusion of the
many different physical aspects of the complicated problem. Figure 11 shows the $\lambda_{model}$
values given as input by the above authors to their detailed formation codes, with the y-axis showing
$\lambda$ from equation (\ref{LamObs}) applied to the final results of the code evolution which followed
the model for $\sim 13$ Gyr. Stars for the van den Bosch (2000) results and triangles for the 
Hernandez et al.(2001) models.

It is interesting that the particular features of a galactic formation scenario
which we are interested in treating here, $\lambda$, final resulting $R_{d}$ and flat rotation region $V_{d}$,
are evidently very well modeled by the trivial physics that went into eq(\ref{LamObs}), across the more than
2.5 orders of magnitude covered by the masses in the modeled galaxies. This can be understood by considering
that eq(\ref{LamObs}) is the result of two fundamental hypothesis: I) an exponentially decreasing surface
density profile for the disk, and II) a dominant dark halo responsible for a flat rotation curve linked
to the disk through a Tully-Fisher relation. Whenever
these two conditions are met, as is the case for the final state of all published modeled galaxies and
real observed systems, basic physics strongly constrains results to lie not far from equation (\ref{LamObs}).
For example, Kregel et al. (2005) clearly identifies the good agreement between the models of 
Dalcanton et al. (1997) and their dynamical and photometric data as an unavoidable corollary of the 
two hypothesis listed above within an approximately
constant $Q$ parameter, irrespective of the actual galactic formation scenario.

This of course,
is not to say that eq(\ref{LamObs}) replaces a detailed galactic formation scenario, which is really the
only way of treating the time evolution of the problem, and to arrive at the physical origin of many of the
well known galactic features which were introduced as empirical facts into the development leading to
 eq(\ref{LamObs}). To mention only two such problems, the origin of Tully-Fisher relation can be traced through
CDM galactic formation simulations (e.g. Avila-Reese et al. 1998, Steinmetz \& Navarro 1999, Navarro \& Steinmetz 2000), 
as can the causes of the observed exponential density surface 
brightness profiles of galaxies (e.g. Saio \& Yoshii 1990, Struck-Marcell 1991, Firmani et al. 1996, Silk 2001, 
Zhang 2000, Bell 2002).

To conclude, we have presented an easily applicable way of estimating the $\lambda$ parameter, a theorist's
favourite descriptor of a galaxy type, for any real observed disk galaxy. We have shown that four 
determinant galactic properties crucial in the
subjective assignment of a Hubble type to an observed disk system correlate well with the introduced
$\lambda$ parameter, in ways very similar to what is seen when comparing against the classical Hubble type
for late type galaxies. We suggest that any other dimensionless galactic parameter for late type disks will
also show analogous trends with $\lambda$ of eq(\ref{LamObs}), as would be expected from Buckinham's theorem
of dimensional analysis (Sedov 1993). The well defined and objective nature of the dimensionless $\lambda$
parameter, both in simulations and in real galaxies through eq(\ref{LamObs}), should make it a useful
tool in comparing the output of numerical galactic formation scenarios and real galaxy samples.

\section*{Acknowledgments}

The authors wish to thank Juliane Dalcanton for her help during the preparation of the manuscript
and the input of an anonymous referee which improved the final version.
The work of X. Hernandez was partly supported by DGAPA-UNAM grant No IN117803-3 and CONACYT grants 
42809/A-1 and 42748. The work of B. Cervantes-Sodi is supported by a CONACYT scholarship.

\end{document}